\documentclass[aip,reprint]{revtex4-1}

\usepackage{graphicx}

\draft % marks overfull lines with a black rule on the right

\begin{document}

%\preprint{}

\title{Extreme Ultraviolet Time- and Angle-Resolved Photoemission Spectroscopy with 21.5 meV Resolution using High-Order Harmonic Generation from a Turn-Key Yb:KGW Amplifier} %Title of paper



\author{Yangyang Liu}
%\thanks{}
\affiliation{Department of Physics, University of Central Florida, Orlando FL 32816, USA}

\author{John E. Beetar}
%\thanks{}
\affiliation{Department of Physics, University of Central Florida, Orlando FL 32816, USA}

\author{Md Mofazzel Hosen}
%\thanks{}
\affiliation{Department of Physics, University of Central Florida, Orlando FL 32816, USA}

\author{Gyanendra Dhakal}
%\thanks{}
\affiliation{Department of Physics, University of Central Florida, Orlando FL 32816, USA}

\author{Christopher Sims}
%\thanks{}
\affiliation{Department of Physics, University of Central Florida, Orlando FL 32816, USA}

\author{Firoza Kabir}
%\thanks{}
\affiliation{Department of Physics, University of Central Florida, Orlando FL 32816, USA}

\author{Marc B. Etienne}
%\thanks{}
\affiliation{Department of Physics, University of Central Florida, Orlando FL 32816, USA}

\author{Klauss Dimitri}
%\thanks{}
\affiliation{Department of Physics, University of Central Florida, Orlando FL 32816, USA}

\author{Sabin Regmi}
%\thanks{}
\affiliation{Department of Physics, University of Central Florida, Orlando FL 32816, USA}

\author{Yong Liu}
%\thanks{}
\affiliation{Ames Laboratory, U.S. Department of Energy, Ames, IA, 50011-3020, USA}

\author{Arjun K. Pathak}
%\thanks{}
\affiliation{Ames Laboratory, U.S. Department of Energy, Ames, IA, 50011-3020, USA}
\affiliation{Department of Physics, SUNY Buffalo State, Buffalo, New York 14222, USA}

\author{Dariusz Kaczorowski}
%\thanks{}
\affiliation{Institute of Low Temperature and Structure Research, Polish Academy of Sciences, PL-50-950 Wroclaw, Poland}

\author{Madhab Neupane}
\email[]{Madhab.Neupane@ucf.edu}
%\thanks{}
\affiliation{Department of Physics, University of Central Florida, Orlando FL 32816, USA}

\author{Michael Chini}
\email[]{Michael.Chini@ucf.edu}
%\thanks{}
\affiliation{Department of Physics, University of Central Florida, Orlando FL 32816, USA}
\affiliation{CREOL, The College of Optics and Photonics, University of Central Florida, Orlando FL 32816, USA}


\date{\today}

\begin{abstract}
%abstract goes here
Characterizing and controlling electronic properties of quantum materials require direct measurements of non-equilibrium electronic band structures over large regions of momentum space. Here, we demonstrate an experimental apparatus for time- and angle-resolved photoemission spectroscopy using high-order harmonic probe pulses generated by a robust, moderately high power ($20 \; \mathrm{W}$) Yb:KGW amplifier with tunable repetition rate between $50$ and $150 \; \mathrm{kHz}$. By driving high-order harmonic generation (HHG) with the second harmonic of the fundamental $1025 \; \mathrm{nm}$ laser pulses, we show that single-harmonic probe pulses at $21.8 \; \mathrm{eV}$ photon energy can be effectively isolated without the use of a monochromator. The on-target photon flux can reach $5 \times 10^{10} \; \mathrm{photons/second}$ at $50 \; \mathrm{kHz}$, and the time resolution is measured to be $320 \; \mathrm{fs}$. The relatively long pulse duration of the Yb-driven HHG source allows us to reach an excellent energy resolution of $21.5 \; \mathrm{meV}$, which is achieved by suppressing the space-charge broadening using a low photon flux of $1.5 \times 10^{8} \; \mathrm{photons/second}$  at a higher repetition rate of $150 \; \mathrm{kHz}$. The capabilities of the setup are demonstrated through measurements in the topological semimetal $\mathrm{ZrSiS}$ and the topological insulator $\mathrm{Sb_{2-x}Gd_{x}Te_{3}}$.
\end{abstract}


\maketitle %\maketitle must follow title, authors, abstract and \pacs


\section{Introduction}
Angle-resolved photoemission spectroscopy (ARPES) is a powerful tool for studying quantum materials, since it can directly detect the electronic band structure and population of the occupied states in complex materials \cite{Damascelli-RMP}. This is achieved experimentally by measuring both the kinetic energy and emission angle of photoelectrons, which can be related to the electron binding energy and the in-plane crystal momentum through energy and momentum conservation laws. However, conventional ARPES has some limitations, for example that it can measure only the occupied electronic states below the Fermi level. Since many fascinating features of quantum materials are hidden in the unoccupied and non-equilibrium states, spectroscopic techniques with energy, momentum, and time resolution are required.

The rapid development of ultrafast laser technology\cite{Brabec-RMP} has enabled novel spectroscopic techniques to access non-equilibrium dynamics in solids on time scales ranging from picoseconds down to attoseconds \cite{Kruchinin-RMP}, which naturally correspond to the time scales of lattice and charge dynamics in materials \cite{Basov-NatMater}. Time-resolved spectroscopies have therefore been instrumental in classifying the mechanisms underlying exotic quantum states, as well as studying the relaxation dynamics of highly-excited quasiparticles. In particular, the development of time- and angle-resolved photoemission spectroscopy (trARPES), which is achieved by combining ultrafast pump-probe spectroscopy with ARPES, can directly measure the non-equilibrium band structure of materials resulting from ultrafast excitation. This technique thereby enables measurement of electron dynamics on femtosecond (fs) to picosecond (ps) timescales in a wide variety of quantum materials, including topological insulators \cite{Madhab-PRL, ZX-Shen-PRL, Gedik-PRL}, high-temperature superconductors \cite{Smallwood-science}, and charge density wave insulators\cite{Hellmann-natcommun, Xun-Shi-CDW, Gedik-CDW-NatPhys}, as well as the discovery of exotic dressed states in quantum materials\cite{Gedik-Science, Gedik-NatPhys}.

In trARPES, a pump pulse is employed to excite dynamics in a material and a subsequent probe pulse is used to image the transient electronic structures. To generate the photoemitted electrons, the photon energy of the probe pulse must be larger than the work function of the material, typically $4-5 \; \mathrm{eV}$. Meeting this requirement requires nonlinear upconversion of the fundamental laser frequency, which can be achieved either by perturbative nonlinear optics or non-perturbative high-order harmonic generation (HHG). trARPES was first demonstrated using probe pulses derived from perturbative harmonic generation in nonlinear crystals, such as $\mathrm{BaB_{2}O_{4}}$ (BBO) \cite{BBO-1,BBO-2,BBO-3,BBO-4,BBO-5} or $\mathrm{KBe_{2}BO_{3}F_{2}}$ (KBBF) \cite{kbbf-trARPES-1}. This method has the advantage of using phase matching to achieve high energy and momentum resolution \cite{guodong-liu-ARPES}, comparable to synchrotron-based ARPES setups. However, the photon energy is limited to $\leq 7 \; \mathrm{eV}$, thereby limiting access to large in-plane momenta and preventing access to the full Brillouin zone in most materials. To overcome this, HHG has recently been applied to generate extreme ultraviolet (XUV) probe pulses with photon energy in the range of $10-50 \; \mathrm{eV}$ for trARPES \cite{HHG-trARPES-1, HHG-trARPES-2, HHG-trARPES-3, HHG-trARPES-4, HHG-trARPES-5, HHG-trARPES-6, HHG-trARPES-7}, enabling access to the full Brillouin zone of many materials. However, the energy resolution achieved in early works with HHG sources was comparatively poor, typically around $90-500 \; \mathrm{meV}$. More recently, energy resolutions of $60 \; \mathrm{meV}$ \cite{HHG-trARPES-5} and $30 \; \mathrm{meV}$ \cite{gedik-NC} have been demonstrated using Ti:Sapphire amplifiers, while cavity-enhanced HHG from Yb fiber lasers can reach an energy resolution of $22 \; \mathrm{meV}$ \cite{Cavity-enhanced}.

Three main factors are responsible for the loss of energy resolution in HHG-based trARPES. First, most setups rely upon Ti:Sapphire amplifiers, with typical output pulse durations of $25$ to $35 \; \mathrm{fs}$, to generate high-order harmonics. Due to the high sensitivity of HHG to the laser field strength, the harmonic pulses are even shorter. Assuming a Gaussian pulse shape, a $20 \; \mathrm{fs}$ harmonic pulse would result in a full-width at half-maximum spectral bandwidth of $90 \; \mathrm{meV}$. To improve the spectral resolution beyond this level, longer driving laser pulses are needed \cite{YbKGW-HHG}. A second limiting factor is the vacuum space charge effect \cite{BBO-2, BBO-4, HHG-trARPES-5, space-charge-Zhou, space-charge-Bauer, space-charge-Lanzara}. Space charge broadening occurs due to Coulomb repulsion of electrons emitted from the sample surface, and can lead to significant ($>1 \; \mathrm{eV}$) energy shifts in the photoelectron spectrum and severe degradation of the energy resolution. The effect can be mitigated by driving HHG with high repetition rate, low-energy laser pulses, for example in an enhancement cavity \cite{HHG-trARPES-2}. However, in most cases, one must make compromises to simultaneously optimize the XUV flux, signal-to-noise ratio, and energy resolution for a particular experiment. Typically, the best energy resolution can be obtained only by using a monochromator, both to select the single harmonic order \cite{monochromator-Dakovski} and to narrow the XUV linewidth \cite{gedik-NC}. However, it has recently been shown that the monochromator can be avoided when driving HHG with short wavelength lasers, since the spectral separation between the neighboring odd harmonics increases with the fundamental laser frequency. For example, by driving HHG with the second harmonic ($\lambda_{L} \approx 400 \; \mathrm{nm}$) of a Ti:Sapphire laser, aluminum and tin filters can be used to select a single harmonic order\cite{HHG-trARPES-1, HHG-trARPES-5, HeWang-NatCommun}. This approach has the additional benefits of a narrower harmonic spectrum\cite{HeWang-NatCommun} and higher efficiency of HHG\cite{400HHG}.

Here, we present a novel setup for trARPES based on a moderately high power ($20 \; \mathrm{W}$), high repetition rate (variable from $50$ to $150 \; \mathrm{kHz}$) Yb:KGW laser. We generate high-order harmonics from the second harmonic ($\lambda_{SH} \approx 511 \; \mathrm{nm}$) of the fundamental $1025 \; \mathrm{nm}$ laser pulses, thereby enabling single harmonic selection using foil filters, and take advantage of the long pulse duration ($280 \; \mathrm{fs}$) to generate narrow-bandwidth probe pulses. By tuning the laser repetition rate, we minimize the effects of space charge broadening while maintaining a relatively high (${>} 10^{8} \; \mathrm{photons/second}$) average harmonic flux. We obtain an energy resolution of $21.5 \; \mathrm{meV}$, allowing us to resolve fine features in the photoelectron spectrum, and a time resolution of $320 \; \mathrm{fs}$. We demonstrate the performance of the setup through time-resolved measurements of the non-equilibrium electronic band structure of $\mathrm{ZrSiS}$, a new topological semimetal with Dirac-like surface states near the edge of the Brillouin zone. The setup has the additional advantage of being highly flexible, with the potential for improvement of the time resolution to below $15 \; \mathrm{fs}$ level through nonlinear spectral broadening\cite{Beetar-JOSAB} and the capability to pass multiple harmonics for attosecond experiments\cite{attoARPES-KM, attoARPES-UKeller}.

\section{Experimental Setup}
\subsection{Vacuum Beamline}

\begin{figure*}[b]
    \includegraphics[width=6.69in,angle=0]{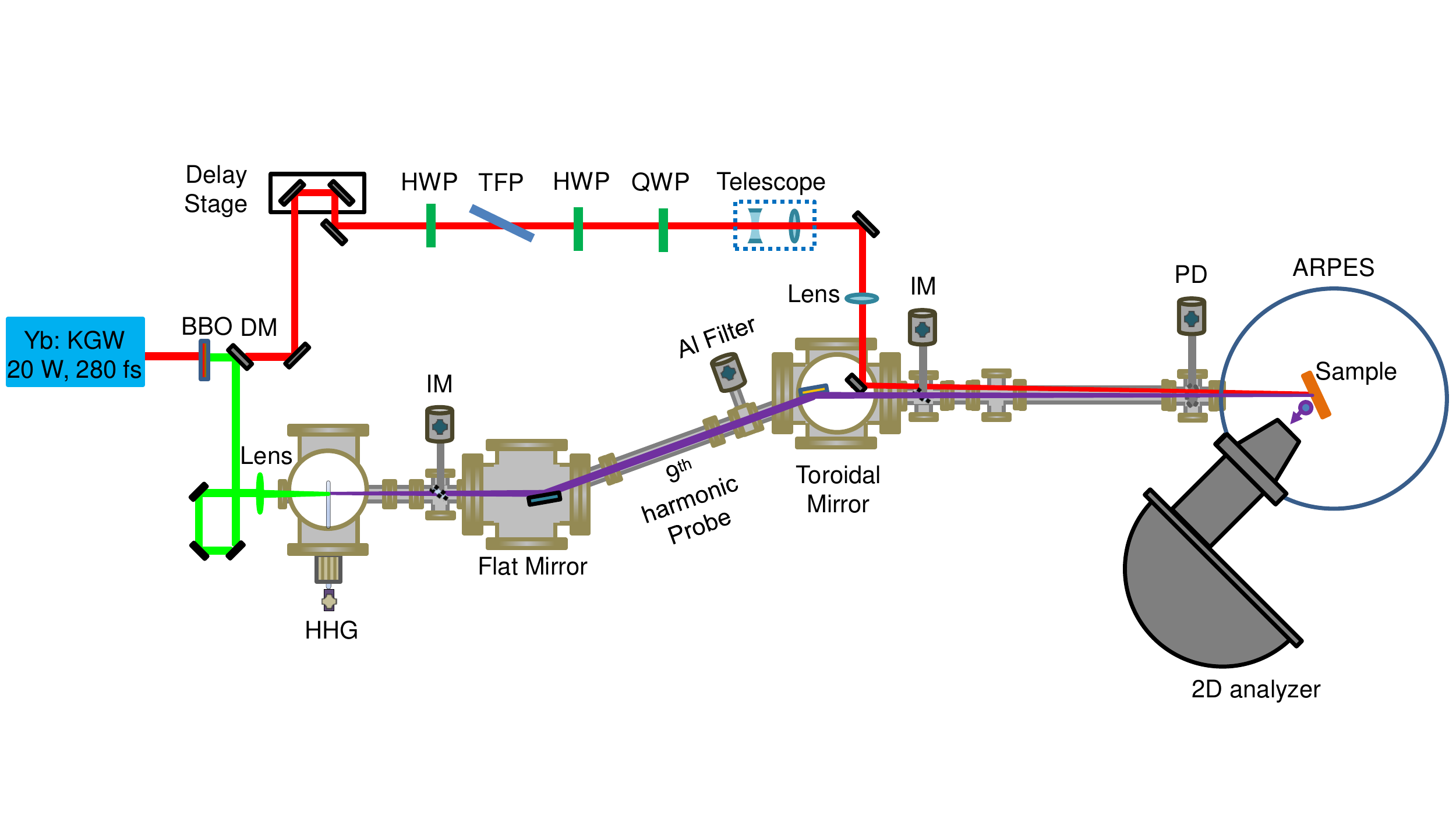}
	\caption{Schematic of experimental setup for 		trARPES. HWP: half-wave plate, TFP: thin-film polarizer, QWP: quarter-wave plate, DM: dichroic mirror, IM: inserting mirror, PD: photodiode.}
	\label{Fig1}
\end{figure*}

The trARPES setup is schematically illustrated in Fig. \ref{Fig1}. The pump and probe pulses are both derived from a commercial, turn-key Yb:KGW amplifier (Light Conversion PHAROS), which emits $280 \; \mathrm{fs}$ pulses with central wavelength $\lambda_{L} = 1025 \; \mathrm{nm}$ and average power of $20 \; \mathrm{W}$ for repetition rates between $50$ and $200 \; \mathrm{kHz}$. For the experiments presented here, repetition rates in the range of $50$ ($400 \; \mathrm{\mu J}$ pulse energy) to $150 \; \mathrm{kHz}$ ($133 \; \mathrm{\mu J}$ pulse energy) are used. Due to the increased efficiency and energy separation between the neighboring harmonics, we choose to drive HHG with the second harmonic of the laser output. Second harmonic pulses centered at $511 \; \mathrm{nm}$ are generated with more than $50 \%$ efficiency using a $2 \; \mathrm{mm}$ thick BBO crystal. The pulse duration of the second harmonic pulses is estimated to be $200 \; \mathrm{fs}$ from the sum-frequency cross-correlation with the fundamental pulses. After generating the second harmonic, a dichroic mirror is used to split the fundamental and second harmonic pulses, respectively, into the pump and probe arms of a Mach-Zehnder interferometer. In the probe arm, high-order harmonics are generated by focusing the second harmonic pulses ($f=175 \; \mathrm{mm}$) into a rectangular capillary (Wale Apparatus, $0.4 \times 4 \; \mathrm{mm}$ inner diameter) backed with krypton gas. The gas cell is installed on a three-axis manipulator to allow alignment of the laser through the $\sim 50 \; \mathrm{\mu m}$ laser-drilled entrance and exit holes in the gas cell and optimization of the phase matching conditions for efficient on-axis harmonics.

After being reflected by a flat mirror, the XUV pulses are focused by a toroidal mirror ($f = 650 \; \mathrm{mm}$ with grazing incidence angle of $10 \; \mathrm{degrees}$, ARW Optical) onto the sample. Both mirrors are coated with $74.5 \; \mathrm{nm}$ silicon carbide (SiC, coated by NTT-AT), leading to overall transmission rates of $62.9 \%$ and $0.3 \%$ for the 9th harmonic and the $511 \; \mathrm{nm}$ pulses, respectively. The distance between the gas cell and the sample is $2.6 \; \mathrm{m}$, with the toroidal mirror imaging the harmonic source in a $4f$ geometry. Between the flat mirror and the toroidal mirror, a $0.5 \; \mathrm{\mu m}$ thick Al filter (Lebow Company) is used to block the residual $511 \; \mathrm{nm}$ pulses and lower order harmonics. Movable mirrors can be inserted after the HHG chamber and after the toroidal mirror, in order to monitor the transmission of the 511 nm through the gas cell and to check the alignment of the toroidal mirror. Close to the entrance to the ARPES chamber, a photodiode (Opto Diode AXUV100AL) is employed to measure the XUV photon flux. The XUV focal spot size is estimated by imaging the fluorescence of a Ce:YAG crystal mounted on the sample plate, yielding a FWHM spot size of $95 \; \mathrm{\mu m} \times 125 \; \mathrm{\mu m}$. 

The energy and momentum of photoelectrons are measured using a high resolution hemispherical analyzer (Scienta Omicron R$3000$). The energy resolution $\Delta E$ of the analyzer can be approximated as $\Delta E \approx \frac{s E_{p}}{2r}$, where $s$ is the slit width, $E_{p}$ is the pass energy and $r$ is the mean radius of the analyzer. For the slit width of $0.2 \; \mathrm{mm}$, the analyzer resolution can reach below $3 \; \mathrm{meV}$ when setting the pass energy to $2 \; \mathrm{eV}$. The angular resolution of the analyzer is $0.1 \; \mathrm{degree}$. Samples are mounted on an XYZ manipulator with primary and azimuthal rotations, and cooled using a liquid helium cryostat. In addition to the laser source, the analyzer is equipped with a helium discharge lamp ($h \nu = 21.2$ or $40.8 \; \mathrm{eV}$) for static measurements.

The pump pulses are derived from the residual $1025 \; \mathrm{nm}$ pulses which transmit through a dichroic mirror placed after the BBO crystal. A delay stage (Newport, DL$125$), with a scan range of $0.8 \; \mathrm{ns}$ and minimum step size of $0.5 \; \mathrm{fs}$, is used to control the time delay between the pump and probe pulses. After the delay stage, a half-wave plate and thin-film polarizer are used to vary the pump intensity, and a second half-wave plate and quarter-wave plate are used to vary the polarization state of the pump pulses. After enlarging the beam size using a telescope, the pump pulses are loosely focused by an $f = 1500 \; \mathrm{mm}$ lens and reflected by a D-shaped mirror onto the target location, yielding a beam spot size of $2w = 450 \; \mathrm{\mu m}$ on the sample. The angle between the pump and probe paths is $0.4 \; \mathrm{degrees}$. Spatial and temporal overlap of the pump and probe pulses are found by reflecting the pump beam and the second harmonic driving beam, which serves as a reference for the XUV, out of the chamber using the movable mirror placed after the toroidal mirror chamber.

A major challenge of HHG-based trARPES measurements is the low pressure, typically on the order of $10^{-10} \; \mathrm{torr}$ or better, needed to prevent surface contamination of the samples in the ARPES chamber. With the gate valve at the ARPES entrance closed, the vacuum inside the ARPES chamber can be maintained below $10^{-10} \; \mathrm{torr}$. However, there is no window which can effectively transmit both the XUV and NIR pulses when performing trARPES measurements, and it is therefore necessary to maintain ultrahigh vacuum levels when the gate valve is open and gas is flowing in the HHG chamber. We have installed turbomolecular pumps (Leybold MAG W $600$iP) on both the HHG chamber and the toroidal mirror chamber, as well as an ion pump (Gamma Vacuum TiTan $75$S) between the toroidal mirror chamber and the ARPES chamber. When performing HHG experiments, the pressure in the HHG chamber and toroidal mirror chamber can reach as high as $10^{-4} \; \mathrm{torr}$. Due to the differential pumping enabled by the Al filter, however, we maintain a pressure below $2 \times 10^{-9} \; \mathrm{torr}$ in the toroidal mirror chamber. A copper gasket with a small hole ($10 \; \mathrm{mm}$ diameter) is installed near the entrance to the ARPES chamber, which also aids in the differential pumping and allows us to achieve a pressure below $10^{-10} \; \mathrm{torr}$ inside the ARPES chamber. Under these conditions, we have found that high quality surface band structures can be obtained over an entire day of measurements.

\subsection{XUV Source Characterization}

\begin{figure*}
    \includegraphics[width=6.69in,angle=0]{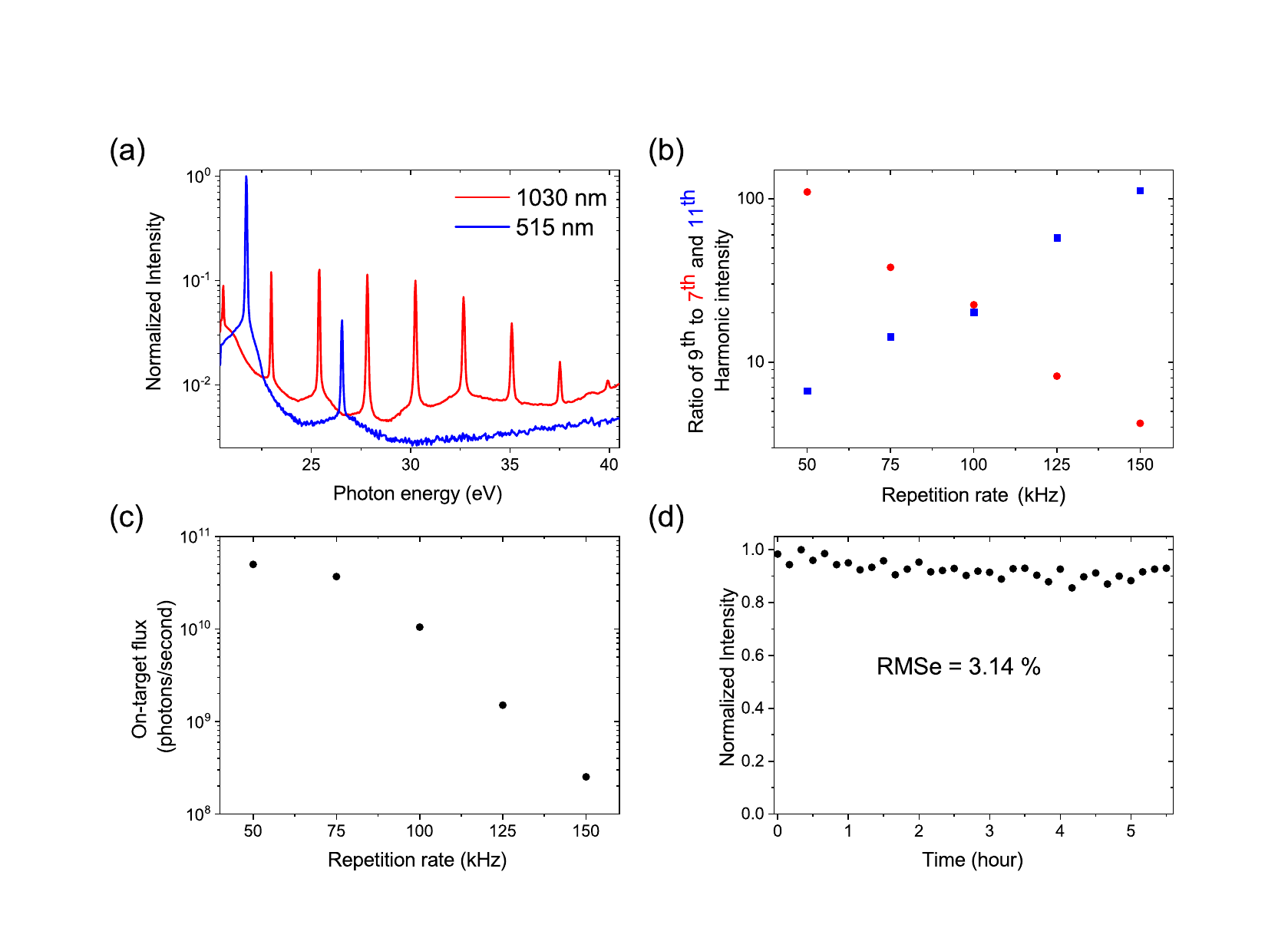}
	\caption{Characterization of XUV high-order harmonic source. (a) Comparison of high-order harmonics generated with the fundamental and second harmonic pulses. (b) Intensity contrast of the 7th, 9th, and 11th harmonic orders. The intensity ratios of the 9th to 7th and 9th to 11th harmonics are shown as red circles and blue squares, respectively. For repetition rates greater than $75 \; \mathrm{kHz}$, the 9th harmonic is more than ten times stronger than the 11th. (c) On-target 9th harmonic flux. At low repetition rates, the average flux approaches $10^{11} \; \mathrm{photons/s}$, and is maintained above $10^{9} \; \mathrm{photons/s}$ for repetition rates of $125 \; \mathrm{kHz}$ and below. (d) Measured stability of the 9th harmonic over a period of $5.5$ hours. The data in (d) was measured by measuring harmonic spectra using the grating spectrometer, with a laser repetition rate of $50 \; \mathrm{kHz}$ and a camera integration time of $\sim 30 \; \mathrm{ms}$.}
	\label{Fig2}
\end{figure*}

The high-order harmonic spectra are characterized using an identical setup for HHG, which is connected to a home-built extreme ultra-violet (XUV) spectrometer consisting of a dispersive flat-field grating (Hitachi 001-0640) and micro-channel plate and phosphor screen detector (Photonis) \cite{Beetar-JOSAB}. The driving laser pulses are blocked using a $0.5 \; \mathrm{\mu m}$-thick Al filter, which also assists in the isolation of a single harmonic order by blocking harmonics below the 9th order. The measured transmission of the filter is $2.5 \%$ for the 9th order and less than $0.1 \%$ for the 7th order. Higher harmonics are not efficiently generated in our experiments, as the relatively low driving laser intensity limits the harmonic cutoff photon energy. The resulting spectrum consists of a strong 9th harmonic peak at $21.8 \; \mathrm{eV}$, with weak peaks at $\sim 17 \; \mathrm{eV}$ and $26.7 \; \mathrm{eV}$, corresponding to the 7th and 11th harmonic orders, respectively. Figure \ref{Fig2}(a) shows a comparison of the harmonic spectra generated with the fundamental $1025 \; \mathrm{nm}$ pulses and the second harmonic pulses. The isolation of the 9th harmonic is displayed in Fig. \ref{Fig2}(b), from which we can see the 9th order is more than 10 times stronger than the 7th and 11th orders for repetition rates between $50$ and $125 \; \mathrm{kHz}$. At $150 \; \mathrm{kHz}$, the 9th harmonic is around $4$ times stronger than the 7th. While the 7th harmonic is not insignificant at high repetition rates, we note that the large energy separation ($4.8 \; \mathrm{eV}$) between harmonic orders is much larger than the pump photon energy ($1.2 \; \mathrm{eV}$), and that the 7th harmonic will therefore have no influence on the photoelectrons close to the Fermi level. Improved isolation of the 9th harmonic could be achieved using other filters, for example a combination of Sn and Ge. The full-width at half-maximum spectral bandwidth of the 9th harmonic is measured to be below $70 \; \mathrm{meV}$. We note, however, that this measurement is limited by the spectral resolution of the grating spectrometer, which is mainly determined by the micro-channel plate and phosphor screen detector \cite{Xiaowei-Wang-AO}. Resolution measurements based on the photoelectron spectrum will be discussed in more detail below.

trARPES measurements are intrinsically multi-dimensional, as energy-momentum spectra must be collected along different symmetry axes of the crystalline sample and for a wide range of pump-probe delays. Therefore, a high average photon flux is necessary to reduce the data collection time, while maintaining high signal-to-noise ratio. However, it is also necessary to limit the single-shot photon flux, which is responsible for space-charge broadening. We attempt to maintain a moderately high average photon flux (${>} 10^{8} \; \mathrm{photons/second}$) with a low single-shot photon flux (${<} 10^{4} \; \mathrm{photons/shot}$) through increasing the repetition rate of the driving laser pulses. First, we optimize the phase matching of HHG\cite{Cord-Arnold-JPB, Mette-Gaarde-JPB, Eric-constant-OL} by scanning the gas cell position relative to the focal spot and optimizing the krypton backing pressure. We find that the single-shot flux of the 9th harmonic is optimized at a repetition rate of $50 \; \mathrm{kHz}$ for a krypton pressure of $20 \; \mathrm{torr}$. Under these conditions, the average photon flux of the 9th order is measured to be $\sim 2 \times 10^{12} \; \mathrm{photons/second}$ using the XUV photodiode. After passing through the $0.5 \; \mathrm{\mu m}$ thick Al filter, with a measured transmission rate of $2.5 \%$, the on-target photon flux is measured to be $5 \times 10^{10} \; \mathrm{photons/second}$, as shown in Fig. \ref{Fig2}(c). This corresponds to $10^{6} \; \mathrm{photons/shot}$. Such a high single-shot flux results in significant space-charge broadening, which will decrease the energy resolution. However, by taking advantage of the tunable repetition rate and per pulse energy, the single-shot flux can be reduced while simultaneously increasing the repetition rate. We obtain an on-target photon flux of $\sim 1 \times 10^{10} \; \mathrm{photons/second}$ ($\sim 10^{5} \; \mathrm{photons/shot}$) when running at $100 \; \mathrm{kHz}$ and $2.5 \times 10^{8} \; \mathrm{photons/second}$ ($\sim 1700 \; \mathrm{photons/shot}$) at $150 \; \mathrm{kHz}$. The stability of the harmonic flux for $50 \; \mathrm{kHz}$ repetition rate is shown in Fig. \ref{Fig2}(d). After an initial warm-up period, the flux is stable for the entire $5.5 \; \mathrm{hour}$ measurement time, with a normalized root-mean-square deviation of $3.14 \%$, which is sufficient to support the pump-probe measurements. 

\section{Time-Resolved Measurements in Topological Materials}
We now turn to the application of our setup to study time-resolved electronic structures in the topological semimetal $\mathrm{ZrSiS}$ and the topological insulator $\mathrm{Sb_{2-x}Gd_{x}Te_{3}}$. We use these measurements to characterize the time and energy resolution of the setup, including the contributions of space-charge broadening to the energy resolution.

\subsection{Topological Semimetal: $\mathrm{ZrSiS}$}

\begin{figure*}
    \includegraphics[width=6.69in,angle=0]{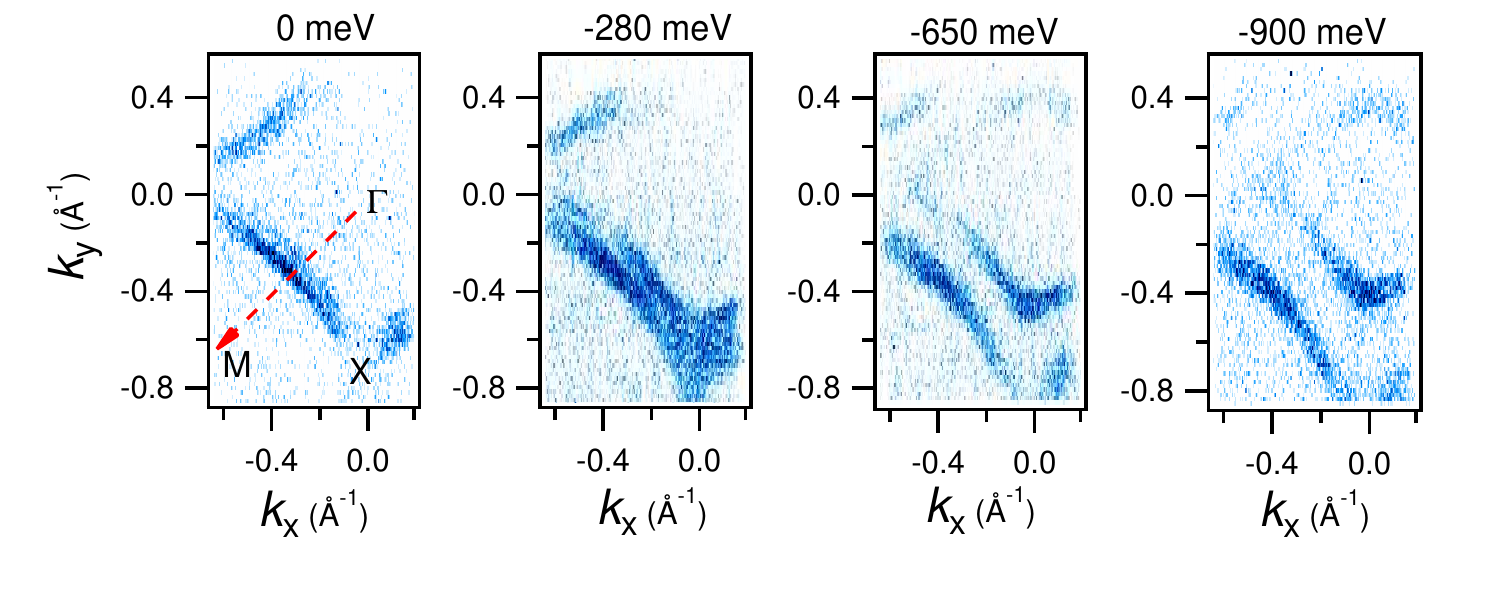}
	\caption{Maps of the electronic band structure in $\mathrm{ZrSiS}$. The figure shows the Fermi surface and constant energy contour plots at various binding energies, which are noted in the plot. The high symmetry points and the nodal line direction (dashed arrow line along the $\mathrm{\Gamma-M}$) are indicated in the leftmost panel. The laser repetition rate we use is 120 kHz.}
	\label{Fig3}
\end{figure*}

\begin{figure*}
    \includegraphics[width=3.37in,angle=0]{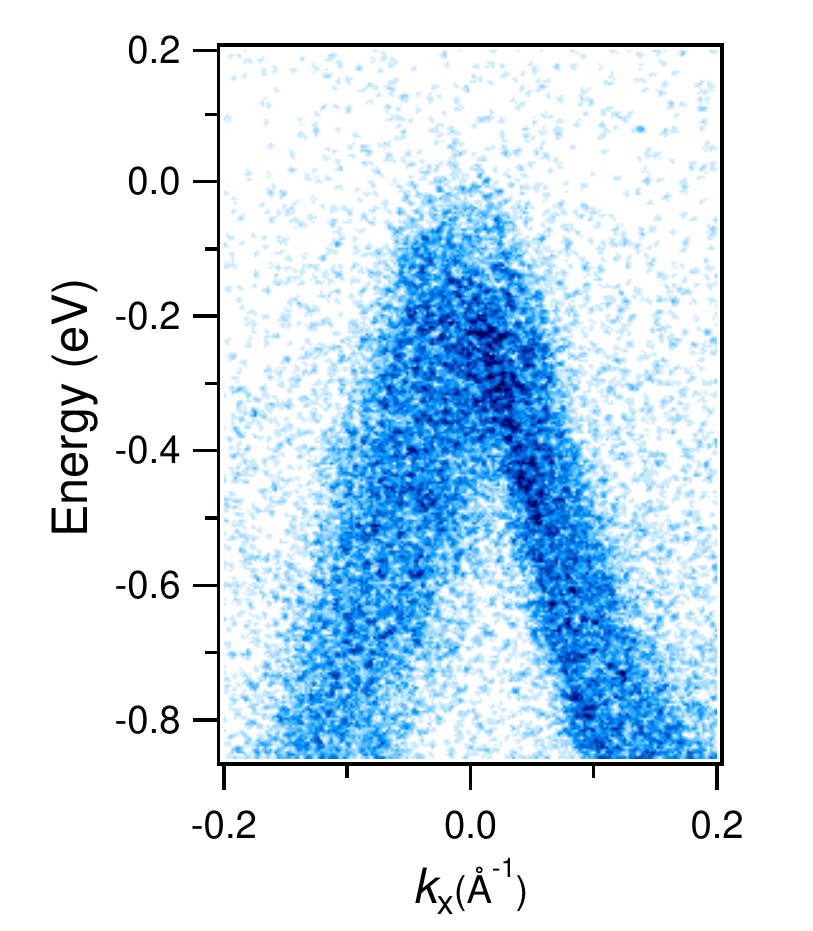}
	\caption{Band dispersion along the nodal line state ($\mathrm{\Gamma-M}$ direction) in $\mathrm{ZrSiS}$, showing linearly-dispersing bands leading to the Dirac point near the Fermi level. The integration time for the measurement was 2 minutes, using a laser repetition rate of 130 kHz.}
	\label{Fig4}
\end{figure*}

In the following, we present preliminary measurements of the topological semimetal $\mathrm{ZrSiS}$. ZrSiS is a novel topological semimetal that exhibits a nodal-line phase which is well separated from any unwanted bulk bands as well as multiple Dirac cones \cite{Madhab-PRB}. Importantly, the Dirac bands show a linear dispersion nature over a wide range of energy (around 2 eV) in this family of materials \cite{Madhab-PRB, ZrSiX-Mofazzel}.We choose $\mathrm{ZrSiS}$ to demonstrate the capabilities of the HHG-based trARPES setup due to the characteristic linearly dispersive states located near the edge of the Brillouin zone. In comparison with laser-based trARPES setups \cite{BBO-1,BBO-2,BBO-3,BBO-4,BBO-5, kbbf-trARPES-1}, the high photon energy available from HHG allows us to access both higher binding energies and larger transverse momentum. This is due to the fact that the parallel momentum can be approximated as $k_{\parallel} = \sqrt{2m_{e}E_{kin}}\sin \theta / \hbar$, where $m_{e}$ is the electron mass, $E_{kin}$ and $\theta$ are the kinetic energy and emission angle of the emitted photoelectron, respectively, and $\hbar$ is the reduced Planck's constant. For a fixed emission angle, higher kinetic energy results in a larger momentum. In our case, with an angular acceptance of $25 \; \mathrm{degrees}$, and electron kinetic energy of $17.3 \; \mathrm{eV}$, the range of momentum space can reach $0.9 \; \mathrm{\AA^{-1}}$. Figure \ref{Fig3} shows the Fermi surface and several energy contours of ZrSiS. The map is composed by measuring $81$ ARPES spectra  for different rotational angles within the range of $-15 \deg$ to $25 \deg$. Each cut was collected with an integration time of $2.5 \; \mathrm{minutes}$, for a total collection time of $203 \; \mathrm{minutes}$. Owing to the relatively large momenta which could be accessed, the map covers most of the Brillouin zone, allowing access to the high-symmetry points. The map, which shows the splitting of the nodal-line state into two separate lines with increasing binding energy, is consistent with synchrotron measurements collected at a higher photon energy of $50 \; \mathrm{eV}$ \cite{Madhab-PRB}. The band structure of the nodal line state is shown in Fig. \ref{Fig4}.

\begin{figure*}
    \includegraphics[width=6.69in,angle=0]{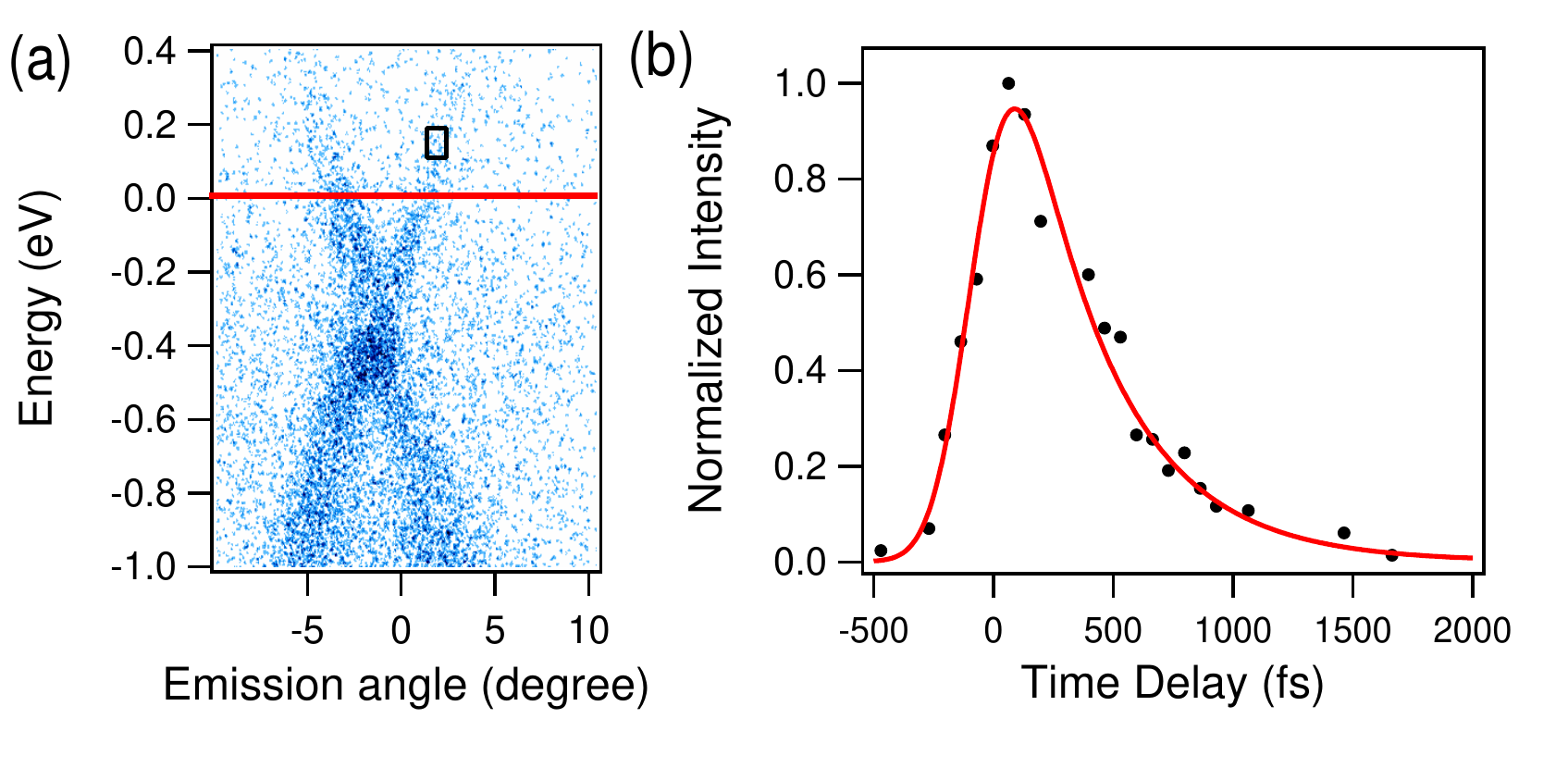}
	\caption{Time-resolved measurements in ZrSiS. (a) Band dispersion of the surface state (close to $\mathrm{X-M}$ direction), showing linearly-dispersing bands with Dirac point below the Fermi level. The extension of the bands above the Fermi level (red line) results from excitation of the pump pulse. (b) Integrated signal within the box in panel (a). The data is fit to a single exponential decay, convoluted with a Gaussian temporal broadening of $320 \; \mathrm{fs}$. The band dispersion in panel (a) and each data point in panel (b) were taken with $2$ minutes integration time at a laser repetition rate of 130 kHz.}
	\label{Fig5}
\end{figure*}

For pump-probe measurements in $\mathrm{ZrSiS}$, we use a pump fluence of $760 \; \mathrm{\mu J / cm^{2}}$ and focus our attention on the surface state. Figure \ref{Fig5}(a) shows the surface state cut (close to $\mathrm{X-M}$ direction) with linearly-dispersive valence bands. The addition of the pump pulse allows us to observe states which are unoccupied under equilibrium conditions \cite{Madhab-PRL}. In Fig. \ref{Fig5}(b), we show the intensity, integrated within a region of energy-momentum space which is accessible only with the pump pulse, as a function of the time delay between the pump and probe. 

\subsection{Time and Energy Resolution}
The source characterization, described in Section IIB above, can provide only crude estimates of the time and energy resolution of the trARPES setup. Based on the measured pulse durations of the $1025 \; \mathrm{nm}$ ($280 \; \mathrm{fs}$) and $511 \; \mathrm{nm}$ ($200 \; \mathrm{fs}$) pulses, we can obtain an upper limit on the time resolution of $340 \; \mathrm{fs}$ and a lower limit on the energy resolution of ~$10 \; \mathrm{meV}$. However, as we expect the XUV pulses to be shorter than those of the $511 \; \mathrm{nm}$ driving laser, it is necessary to characterize both the time and energy resolution from photoelectron measurements.

The time resolution of the trARPES setup can be estimated from the rise time of the time-resolved measurements shown in Fig. \ref{Fig5}. Briefly, the highest-energy states above the Fermi level can be populated only through a single channel, which is the direct absorption of a pump photon. The high-energy electrons will then decay to lower energy states via scattering. The different timescales of these two processes yield an asymmetric distribution, which encodes both the population and decay timescales. We fit the delay-dependent signal to a convolution between a Gaussian and an exponential decay, shown as the red line in Fig. \ref{Fig5}. The fit yields a Gaussian temporal broadening of $320 \; \mathrm{fs}$, which corresponds to the cross-correlation between our $280 \; \mathrm{fs}$ pump pulses and an XUV pulse with duration of $\sim 150 \; \mathrm{fs}$.

Next, we characterize the energy resolution of the setup. Near the Fermi level, the electronic density of states is governed by the Fermi-Dirac distribution. The ultimate limit to the energy resolution of the measurement is therefore determined by the sample temperature, the XUV harmonic bandwidth, and the detector resolution. However, the energy resolution is further degraded by space-charge broadening, particularly when high photon flux and low repetition rates are used. Here, we analyze the contributions of these factors.

\begin{figure*}
    \includegraphics[width=6.69in,angle=0]{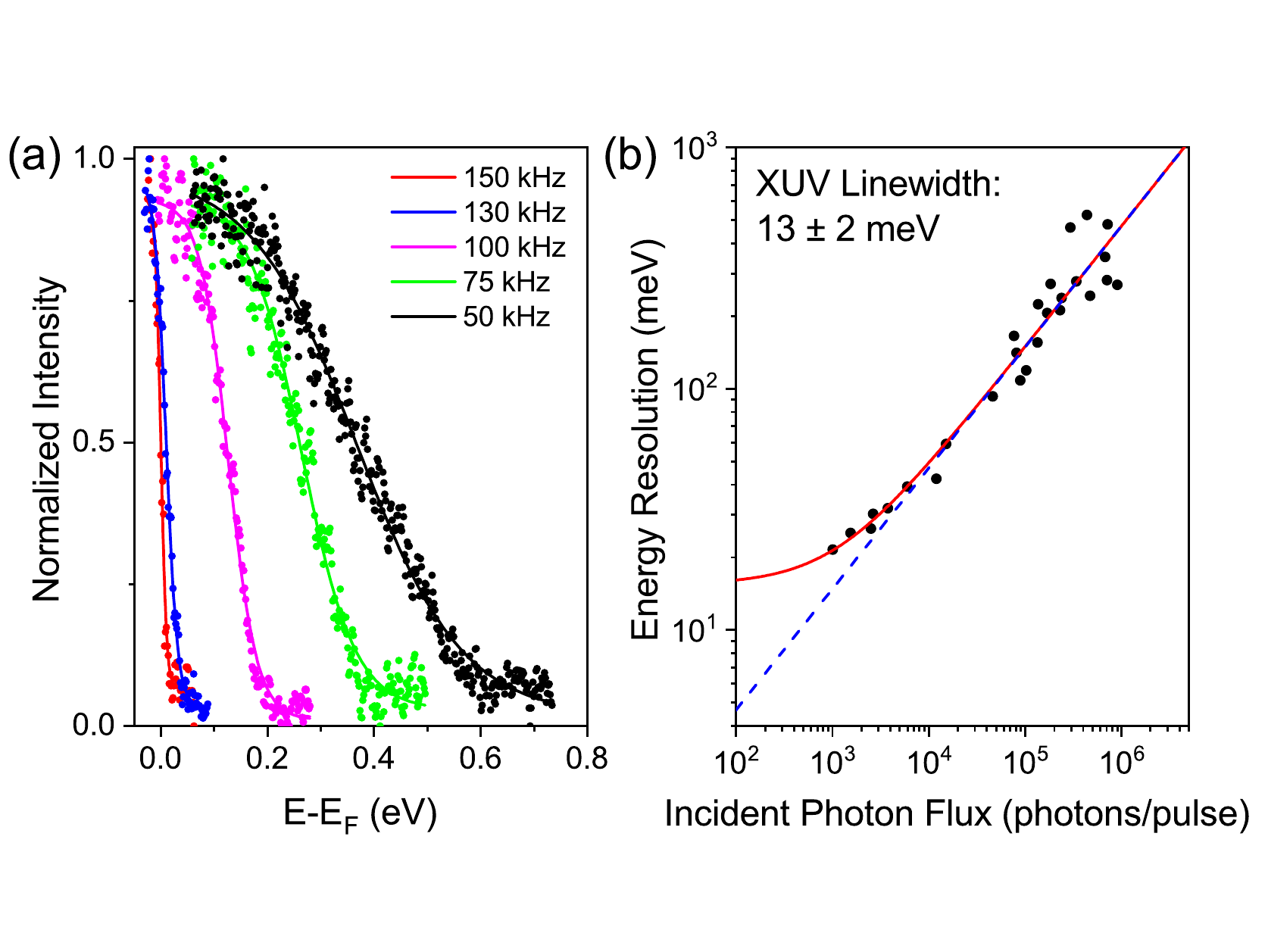}
	\caption{Effects of space-charge broadening and repetition rate on the energy resolution. (a) Momentum-integrated Fermi edge of ZrSiS for different repetition rates at a constant laser power of $20 \; \mathrm{W}$. As the repetition rate is increased, the Fermi level shift and energy broadening of the distribution are minimized. The solid lines show fits to the Fermi-Dirac distribution for each repetition rate. The red dots are taken at a repetition rate of $150 \; \mathrm{kHz}$ and a slightly reduced pulse energy of $\sim 126 \; \mathrm{\mu J}$, and the black dots are taken at $50 \; \mathrm{kHz}$ and reduced pulse energy of $\sim 280 \; \mathrm{\mu J}$. Others are taken with the maximum pulse energies corresponding to an average power of $20 \; \mathrm{W}$ at the listed repetition rate. (b) Dependence of the energy resolution (solid red line) and the space-charge contribution (dashed blue line) on the incident photon flux. At high photon fluxes, the energy resolution follows the square root of the photon flux, while at low photon flux the contributions of other factors become significant.}
	\label{Fig6}
\end{figure*}

\begin{figure*}
    \includegraphics[width=3.37in,angle=0]{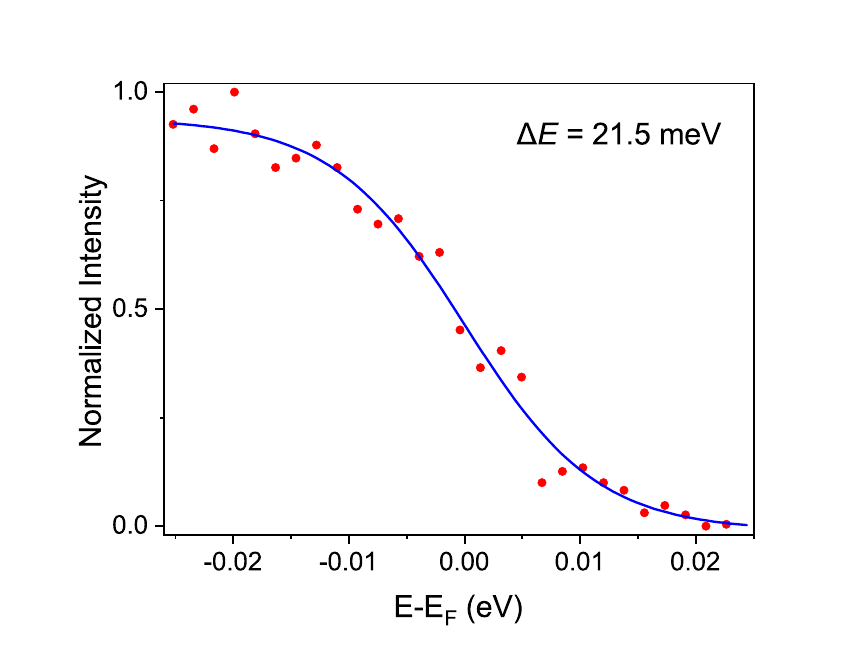}
	\caption{Measurement of the momentum-integrated Fermi edge of ZrSiS at a repetition rate of $150 \; \mathrm{kHz}$ and a slightly reduced pulse energy of $\sim 126 \; \mathrm{\mu J}$, showing a fit to the Fermi-Dirac distribution (blue line) with a temperature of $20 \; \mathrm{K}$ (thermal broadening: $\sim7 \; \mathrm{meV}$) and energy resolution of $21.5 \; \mathrm{meV}$ with an integration time of $1.5$ hours.}
	\label{Fig7}
\end{figure*}

As discussed above, the energy resolution $\Delta E$ is mainly limited by three major factors, approximated as $\Delta E = \sqrt{\Delta E_{XUV}^{2} + \Delta E_{det}^2 + \Delta E_{sc}^{2}}$. The first factor is the XUV spectral bandwidth, $\Delta E_{XUV}$. While we cannot directly measure the spectral bandwidth due to the limited resolution of our XUV spectrometer, we can determine a lower limit for the bandwidth based on the measured time resolution. A transform-limited Gaussian XUV pulse with a FWHM duration of $150 \; \mathrm{fs}$ would yield a spectral bandwidth of approximately $12 \; \mathrm{meV}$. The second factor is the analyzer resolution $\Delta E_{det}$. Under our experimental conditions ($s = 0.2 \; \mathrm{mm}$, $E_{p} = 10 \; \mathrm{eV}$), the analyzer resolution is determined by the manufacturer to be $8.8 \; \mathrm{meV}$. The third and final factor is the space-charge broadening, which depends on the number of photoelectrons generated per pulse. We investigate the scaling of the energy broadening with photon flux, as shown in Fig. \ref{Fig6}, by taking advantage of the tunable repetition rate of our laser. The data shown in Fig. 6 were collected on three different days and for three different ZrSiS samples to ensure repeatability of the measurement. Most of the measurements, and in particular those at the highest repetition rates, were collected for a sample temperature of $\sim 20 \; \mathrm{K}$, while others were collected at $\sim 80 \; \mathrm{K}$. Figure \ref{Fig6}(a) shows the momentum-integrated Fermi edge of ZrSiS for constant average laser power but with varying repetition rate. At low repetition rate (high pulse energy), the energy resolution is dominated by space-charge effects, and significant shifts of the Fermi level on the $>100 \; \mathrm{meV}$ scale can be seen. At high repetition rate (low pulse energy), however, the contributions from the XUV bandwidth and detector resolution, play a major role. We fit the data to a Fermi-Dirac distribution with the measured temperature, and plot the FWHM bandwidth as a function of the per-pulse photon flux $N$, as shown in Fig. \ref{Fig6}(b). For ZrSiS, we observed that the contribution of space-charge broadening scales approximately as $N^{1/2}$, and including this dependence in a fit to the energy broadening allows us to determine the ultimate limits to our resolution. By completely suppressing space-charge effects, we find an ultimate limit to the energy resolution of $16 \; \mathrm{meV}$. In practice, however, achieving such energy resolution is impractical due to the long integration times necessary. When operating the laser at a repetition rate of $150 \; \mathrm{kHz}$ and a slightly reduced pulse energy of $\sim 126 \; \mathrm{\mu J}$, which presents a practical limit for measurements with integration times of 1.5 hours, we find that the contribution of space-charge broadening is $14.7 \; \mathrm{meV}$ and that the energy resolution is $21.5 \; \mathrm{meV}$, as shown in Fig. 7. Our measured energy resolution, in combination with the detector energy resolution measured independently using a helium discharge lamp, yield an XUV linewidth of $13 \pm 2  \; \mathrm{meV}$, which represents more than four times reduction in comparison with the harmonic linewidth that can be obtained from Ti:Sapphire lasers without the use of a monochromator \cite{HHG-trARPES-5}. We further note that the energy resolution we obtain is close to that which can be obtained with laser ARPES at much lower photon energies.

\subsection{Topological Insulator: $\mathrm{Sb_{2-x}Gd_{x}Te_{3}}$}

\begin{figure*}
    \includegraphics[width=6.69in,angle=0]{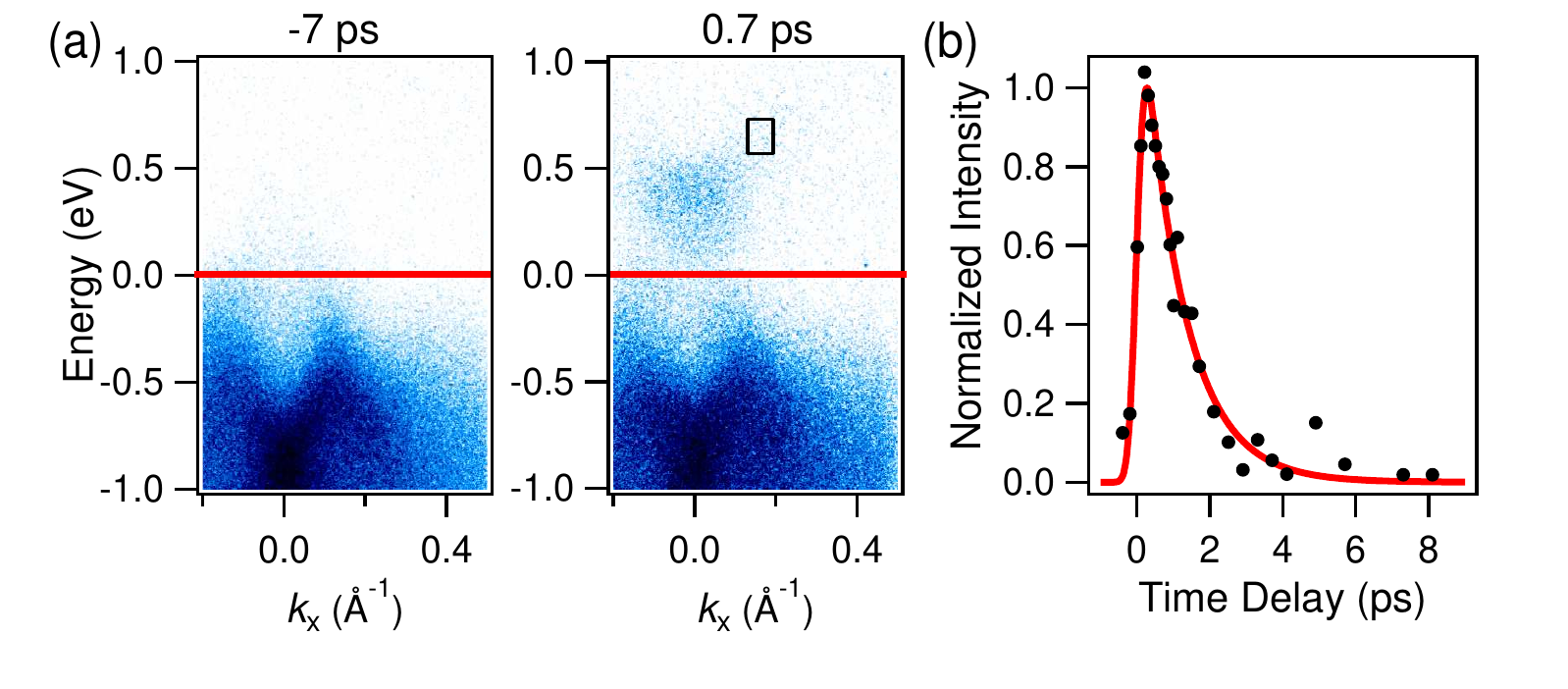}
	\caption{Time-resolved measurements in $\mathrm{Sb_{2-x}Gd_{x}Te_{3}}$. (a) Band dispersion maps along the (${K-\Gamma-K}$ direction). The extension of the bands above the Fermi level (red line) results from excitation of the pump pulse. The time delays are noted on the plots. (b) Integrated signal within the box in panel (a) right. The data fits to an exponentially-modified Gaussian distribution (red line), with a decay lifetime of $1.1 \; \mathrm{ps}$. The band dispersion in panel (a) and each data point in panel (b) were taken with $2$ minutes integration time at a laser repetition rate of $120 \; \mathrm{kHz}$.}
	\label{Fig8}
\end{figure*}

We next show pump-probe measurements on the topological insulator $\mathrm{Sb_{2-x}Gd_{x}Te_{3}}$, with $x=0.01$. $\mathrm{Sb_{2}Te_{3}}$ is one of the well-studied topological insulator systems, and is characterized by a single Dirac cone located at the $\Gamma$ point of the Brillouin zone. It is a hole-doped system, and the Dirac point is therefore located above the Fermi level \cite{TI-SCZ}. Doping with Gd can be used to tune the energy of the Dirac point with respect to the Fermi level in $\mathrm{Sb_{2-x}Gd_{x}Te_{3}}$ as well as to study the effect of magnetism in topological insulator system. The samples are cooled down to $77 \; \mathrm{K}$ and the pump fluence is $2.7\; \mathrm{mJ / cm^{2}}$. Figure \ref{Fig8} shows the band structures measured at pump-probe time delays of $-7 \; \mathrm{ps}$ and $0.7 \; \mathrm{ps}$, as well as the integrated intensity measured as a function of the delay. To analyze the decay lifetime of the excited state, we again fit the integrated signal to an exponentially-modified Gaussian distribution. The decay timescale of $1.1 \; \mathrm{ps}$ is consistent with previous measurements of relaxation dynamics using laser-based trARPES on topological insualtors.

\section{Outlook and Conclusion}
\subsection{Improvement of Time Resolution}
While the setup described above demonstrates a significant improvement in the achievable energy resolution and usable photon flux over Ti:Sapphire-driven HHG-based trARPES systems, it lacks the time resolution needed to classify few-femtosecond dynamics, such as in a quantum phase transition \cite{Hellmann-natcommun, Leone-PNAS}. However, this is not a fundamental limitation. In our laboratory, we have demonstrated the efficient nonlinear compression of our $280 \; \mathrm{fs}$ laser pulses to $15 \; \mathrm{fs}$ in a hollow-core fiber pulse compressor \cite{Beetar-JOSAB}. By tuning the gas pressure in the fiber and optimizing the dispersion compensation, arbitrary pulse durations can in principle be achieved. We estimate, based on the energy throughput of the nonlinear compressor, the decreased efficiency of second harmonic generation for broad bandwiths, and the increased intensity associated with the short pulse duration, that efficient harmonic generation can be maintained for driving pulses of $15 \; \mathrm{fs}$ or shorter.

\subsection{Attosecond Measurements}
Due to the lack of a monochromator, the setup is also well-suited for trARPES measurements with attosecond resolution based on the RABBITT (Reconstruction of Attosecond Beating By Interference of Two-photon Transitions) technique \cite{attoARPES-KM, attoARPES-UKeller}. In RABBITT measurements, multiple harmonics are allowed to hit the sample simultaneously, producing replicas of the momentum-dependent electron energy spectrum separated by twice the fundamental laser photon energy. When the emitted photoelectrons interact with a dressing laser field, sidebands are formed in between the replicas, which oscillate as the relative time delay between the harmonics and the fundamental pulse are varied on attosecond timescales \cite{PM-Paul-science}. These measurements can yield the group delay of the attosecond pulses, as well as the time delay between different photoemission channels \cite{UKeller-Optica}. When narrow-bandwidth harmonics are used, the contributions from closely-spaced channels can be determined while maintaining attosecond time resolution \cite{Huillier-Science}. In the near future, we aim to implement momentum-resolved RABBITT measurements by increasing the driving laser intensity to enable the generation of higher harmonic orders, which can be easily passed by an aluminum foil filter. The second harmonic of the pump pulse will be generated prior to recombination with the XUV, and will serve as the dressing laser pulse for attosecond time-resolved measurements.

\subsection{Conclusion}
We have demonstrated a novel HHG-based setup for time- and angle-resolved photoemission spectroscopy using a commercial $\mathrm{Yb:KGW}$ laser amplifier. The moderately high average power ($20 \; \mathrm{W}$) and high repetition rate (variable from $50$ to $150 \; \mathrm{kHz}$) allow us to generate a moderately high-flux (more than $10^{9} \; \mathrm{photons/s}$ for repetition rates between $50$ and $125 \; \mathrm{kHz}$) in a single XUV harmonic, while mitigating the effects of space-charge broadening. By generating high-order harmonics using the second harmonic of the fundamental $1025 \; \mathrm{nm}$ laser pulses, we isolate the 9th harmonic probe pulse with photon energy of $21.8 \; \mathrm{eV}$ without using a monochromator. By suppressing the space-charge broadening using a low photon flux of $1.5 \times 10^{8} \; \mathrm{photons/second}$, the energy resolution is measured to be $21.5 \; \mathrm{meV}$, allowing us to resolve detailed features in the electronic structure of topological materials. Time-resolved measurements, with temporal resolution of $320 \; \mathrm{fs}$, are demonstrated in the topological semimetal $\mathrm{ZrSiS}$ and topological insulator $\mathrm{Sb_{2-x}Gd_{x}Te_{3}}$. The flexibility of the setup is expected to enable improvement of the time resolution for single-harmonic measurements to around $15 \; \mathrm{fs}$ or to allow passage of multiple harmonics for attosecond spectroscopies.

% If in two-column mode, this environment will change to single-column format so that long equations can be displayed. 
%\begin{widetext}
%$$\mbox{put long equation here}$$
%\end{widetext}

% Figures should be put into the text as floats. 
% Use the graphics or graphicx packages (distributed with LaTeX2e).
% Put the label that you will use with \ref{} command in the braces of the \label{} command.
%
% \begin{figure}
% \end{figure}

% \begin{table}
% \end{tabular}
% \end{table}

\begin{acknowledgments}
This material is based on the work supported by the Air Force Office of Scientific Research under Award Numbers FA9550-16-1-0149 and FA9550-17-1-0415. Work at Ames Laboratory is supported by the U.S. Department of Energy, Office of Basic Energy Sciences, Division of Material Sciences and Engineering under contract No. DE-AC02-07CH11358 with Iowa State University. M. B. Etienne was partially supported by funding from Duke Energy through the UCF EXCEL/COMPASS Undergraduate Research Experience program. M. Chini would like to acknowledge useful discussions with Dr. Thomas Allison.
\end{acknowledgments}

%\bibliography{your-bib-file}

\begin{thebibliography}{xx}% Produces the bibliography via BibTeX.
\bibitem {Damascelli-RMP}A. Damascelli, Z. Hussain, and Z.-X. Shen, Rev. Mod. Phys. {\bfseries 75}, 473 (2003).
\bibitem {Brabec-RMP}T. Brabec, and F. Krausz, Rev. Mod. Phys. {\bfseries 72}, 545 (2000).
\bibitem {Kruchinin-RMP}S. Y. Kruchinin, F. Krausz, and V. S. Yakovlev, Rev. Mod. Phys. {\bfseries 90}, 021002 (2018).
\bibitem {Basov-NatMater}D. N. Basov, R. D. Averitt, and D. Hsieh, Nat. Mater. {\bfseries 16}, 1077 (2017).
\bibitem {Madhab-PRL}M. Neupane, S.-Y. Xu, Y. Ishida, S. Jia, B. M. Fregoso, C. Liu, I. Belopolski, G. Bian, N. Alidoust, T. Durakiewicz, V. Galitski, S. Shin, R. J. Cava, and M. Z. Hasan, Phys. Rev. Lett. {\bfseries 115}, 116801 (2015).
\bibitem {ZX-Shen-PRL} J. A. Sobota, S. Yang, J. G. Analytis, Y. L. Chen, I. R. Fisher, P. S. Kirchmann, and Z.-X. Shen, Phys. Rev. Lett. {\bfseries 108}, 117403 (2012).
\bibitem {Gedik-PRL} Y. H. Wang, D. Hsieh, E. J. Sie, H. Steinberg, D. R. Gardner, Y. S. Lee, P. Jarillo-Herrero, and N. Gedik, Phys. Rev. Lett. {\bfseries 109}, 127401 (2012).
\bibitem {Smallwood-science}C. L. Smallwood, J. P. Hinton, C. Jozwiak, W. Zhang, J. D. Koralek, H. Eisaki, D.-H. Lee, J. Orenstein, and A. Lanzara, Science {\bfseries 336}, 1137 (2012).
\bibitem {Xun-Shi-CDW} X. Shi, W. You, Y. Zhang, Z. Tao, P. M. Oppeneer,
X. Wu, R. Thomale, K. Rossnagel, M. Bauer, H. Kapteyn, M. Murnane, Sci. Adv. {\bfseries 5}, eaav4449 (2019).
\bibitem {Gedik-CDW-NatPhys} A. Zong, A. Kogar, Y. Bie, T. Rohwer, C. Lee, E. Baldini, E. Ergeçen, M. B. Yilmaz, B. Freelon, E. J. Sie, H. Zhou, J. Straquadine, P. Walmsley, P. E. Dolgirev, A. V. Rozhkov, I. R. Fisher, P. Jarillo-Herrero, B. V. Fine, and N. Gedik, Nat. Phys. {\bfseries 15}, 27 (2019).
\bibitem {Hellmann-natcommun}S. Hellmann, T. Rohwer, M. Kalläne, K. Hanff, C. Sohrt, A. Stange, A. Carr, M. M. Murnane, H. C. Kapteyn, L. Kipp, M. Bauer, and K. Rossnagel, Nat. Commun. {\bfseries 3}, 1069 (2012).
\bibitem {Gedik-Science}Y. H. Wang, H. Steinberg, P. Jarillo-Herrero, and N. Gedik, Science {\bfseries 342}, 453 (2013).
\bibitem {Gedik-NatPhys} F. Mahmood, C. Chan, Z. Alpichshev, D. Gardner, Y. Lee, P. A. Lee, and N. Gedik, Nat. Phys. {\bfseries 12}, 306 (2016).
\bibitem {BBO-1} C. L. Smallwood, C. Jozwiak, W. Zhang, and A. Lanzara, Rev. Sci. Instrum. {\bfseries 83}, 123904 (2012).
\bibitem {BBO-2}Y. Ishida, T. Togashi, K. Yamamoto, M. Tanaka, T. Kiss, T. Otsu, Y. Kobayashi, and S. Shin, Rev. Sci. Instrum. {\bfseries 85}, 123904 (2014).
\bibitem {BBO-3}F. Boschini, H. Hedayat, C. Dallera, P. Farinello, C. Manzoni, A. Magrez, H. Berger, G. Cerullo, and E. Carpene, Rev. Sci. Instrum. {\bfseries 85}, 123903 (2014).
\bibitem {BBO-4}J. Faure, J. Mauchain, E. Papalazarou, W. Yan, J. Pinon, M. Marsi, and L. Perfetti, Rev. Sci. Instrum. {\bfseries 83}, 043109 (2012).
\bibitem {BBO-5}E. Carpene, E. Mancini, C. Dallera, G. Ghiringhelli, C. Manzoni, G. Cerullo, and S. D. Silvestri, Rev. Sci. Instrum. {\bfseries 80}, 055101 (2009).
\bibitem {kbbf-trARPES-1} Y. Yang, T. Tang, S. Duan, C. Zhou, D. Hao, and W. Zhang, Rev. Sci. Instrum. {\bfseries 90}, 063905 (2019).
\bibitem {guodong-liu-ARPES} G. Liu, G. Wang, Y. Zhu, H. Zhang, G. Zhang, X. Wang, Y. Zhou, W. Zhang, H. Liu, L. Zhao, J. Meng, X. Dong, C. Chen, Z. Xu, and X. J. Zhou, Rev. Sci. Instrum. {\bfseries 79}, 023105 (2008).
\bibitem {HHG-trARPES-1} S. Eich, A. Stange, A.V. Carr, J. Urbancic, T. Popmintchev, M. Wiesenmayer,K. Jansen, A. Ruffing, S. Jakobs, T. Rohwer, S. Hellmann, C. Chen, P. Matyba,L. Kipp, K. Rossnagel, M. Bauer, M.M. Murnane, H.C. Kapteyn, S. Mathias, and M. Aeschlimann, J. Electron Spectrosc. Relat. Phenom. {\bfseries 195}, 231 (2014).
\bibitem {HHG-trARPES-2} C. Corder, P. Zhao, J. Bakalis, X. Li,
M. D. Kershis, A. R. Muraca, M. G. White, and T. K. Allison, Stru. Dyn. {\bfseries 5}, 054301 (2018).
\bibitem {HHG-trARPES-3}S. Mathias, L. Miaja-Avila, M. M. Murnane, H. Kapteyn, M. Aeschlimann, and M. Bauer, Rev. Sci. Instrum. {\bfseries 78}, 083105 (2007).
\bibitem {HHG-trARPES-4}G. Rohde, A. Hendel, A. Stange, K. Hanff, L.-P. Oloff, L. X. Yang, K. Rossnagel, and M. Bauer, Rev. Sci. Instrum. {\bfseries 87}, 103102 (2016).
\bibitem {HHG-trARPES-5}J. H. Buss, H. Wang, Y. Xu, J. Maklar, F. Joucken ,L. Zeng, S. Stoll, C. Jozwiak, J. Pepper, Y.Chuang , J. D. Denlinger, Z. Hussain, A. Lanzara, and R. A. Kaindl, Rev. Sci. Instrum. {\bfseries 90}, 023105 (2019).
\bibitem {HHG-trARPES-6} R. Wallauer, J. Reimann, N. Armbrust, J. Güdde, and U. Höfer, Appl. Phys. Lett. {\bfseries 109}, 162102 (2016).
\bibitem {HHG-trARPES-7} B. Frietsch, R. Carley, K. Döbrich, C. Gahl, M. Teichmann, O. Schwarzkopf, Ph. Wernet, and M. Weinelt, Rev. Sci. Instrum. {\bfseries 84}, 075106 (2013).
\bibitem {gedik-NC} E. J. Sie, T. Rohwer, C. Lee, and N. Gedik, Nat. Commun. {\bfseries 10}, 3535 (2019).
\bibitem {Cavity-enhanced} A. K. Mills, S. Zhdanovich, M. X. Na, F. Boschini, E. Razzoli, M. Michiardi, A. Sheyerman, M. Schneider, T. J. Hammond, V. Süss, C. Felser, A. Damascelli, and D. J. Jones, Rev. Sci. Instrum. {\bfseries 90}, 083001 (2019).
\bibitem {YbKGW-HHG}E. Lorek, E. W. Larsen, C. M. Heyl, S. Carlström, D. Paleček, D. Zigmantas, and J. Mauritsson, Rev. Sci. Instrum. {\bfseries 85}, 123106 (2014).
\bibitem {space-charge-Zhou} X.J. Zhou, B. Wannberg, W. L. Yang, V. Brouet, Z. Sun, J. F. Douglas, D. Dessau, Z. Hussain, Z.-X. Shen, J. Electron Spectrosc. Relat. Phenom. {\bfseries 142}, 27 (2005).
\bibitem {space-charge-Bauer} S. Passlack, S. Mathias, O. Andreyev, D. Mittnacht, M. Aeschlimann, and M. Bauer, J. Appl. Phys. {\bfseries 100}, 024912 (2006).
\bibitem {space-charge-Lanzara} J. Graf, S. Hellmann, C. Jozwiak, C. L. Smallwood, Z. Hussain, R. A. Kaindl, L. Kipp, K. Rossnage, and A. Lanzara, J. Appl. Phys. {\bfseries 107}, 014912 (2010).
\bibitem {monochromator-Dakovski} G. L. Dakovski, Y. Li, T. Durakiewicz, and G. Rodriguez, Rev. Sci. Instrum. {\bfseries 81}, 073108 (2010).
\bibitem {HeWang-NatCommun} H. Wang, Y. Xu, S. Ulonska, J. S. Robinson, P. Ranitovic, and R. A. Kaindl, Nat. Commun. {\bfseries 6}, 7459 (2015).
\bibitem {400HHG} E. L. Falcão-Filho, C.-J. Lai, K.-H. Hong, V.-M. Gkortsas, S.-W. Huang, L.-J. Chen, and F. X. Kärtner, Appl. Phys. Lett. {\bfseries 97}, 061107 (2010).
\bibitem {Beetar-JOSAB} J. E. Beetar, F. Rivas, S. Gholam-Mirzaei, Y. Liu, and M. Chini, J. Opt. Soc. Am. B. {\bfseries 36}, A33 (2019).
\bibitem {attoARPES-KM} Z. Tao, C. Chen, T. Szilvási, M. Keller, M. Mavrikakis, H. Kapteyn, M. Murnane, Science {\bfseries 353}, 62 (2016).
\bibitem {attoARPES-UKeller} L. Kasmi, M. Lucchini, L. Castiglione, P. Kliuiev, J. Osterwalder, M. Hengsberger, L. Gallmann, P. Krüger, and U. Keller, Optica {\bfseries 4}, 1492 (2017).
\bibitem {Xiaowei-Wang-AO} X. Wang, M. Chini, Y. Cheng, Y. Wu, and Z. Chang, Appl. Opt. {\bfseries 52}, 323 (2013).
\bibitem {Cord-Arnold-JPB} C. M. Heyl, C. L. Arnold, A. Couairon, and A. L’Huillier, J. Phys. B: At. Mol. Opt. Phys. {\bfseries 50}, 013001 (2017).
\bibitem {Mette-Gaarde-JPB} M. B. Gaarde, J. L. Tate, and K. J. Schafer, J. Phys. B: At. Mol. Opt. Phys. {\bfseries 41}, 132001 (2008).
\bibitem {Eric-constant-OL} A. Cabasse, G. Machinet, A. Dubrouil, E. Cormier, and E. Constant, Opt. Lett. {\bfseries 37}, 4618 (2012).
\bibitem {Madhab-PRB} M. Neupane, I. Belopolski, M. M. Hosen, D. S. Sanchez, R. Sankar, M. Szlawska, S.-Y. Xu, K. Dimitri, N. Dhakal, P. Maldonado, P. M. Oppeneer, D. Kaczorowski, F. Chou, M. Z. Hasan, and T. Durakiewicz, Phys. Rev. B {\bfseries 93}, 201104(R) (2010).
\bibitem {ZrSiX-Mofazzel} M. M. Hosen, K. Dimitri, I. Belopolski, P. Maldonado, R. Sankar, N. Dhakal, G. Dhakal, T. Cole, P. M. Oppeneer, D. Kaczorowski, F. Chou, M. Z. Hasan, T. Durakiewicz, and M. Neupane, Phys. Rev. B {\bfseries 95}, 161101(R) (2017).
\bibitem {TI-SCZ}H. Zhang, C.-X. Liu, X.-L. Qi, X. Dai, Z. Fang, and S.-C. Zhang, Nat. Phys. {\bfseries 5}, 438 (2009).
\bibitem {Leone-PNAS} M. F. Jager, C. Ott, P. M. Kraus, C. J. Kaplan, W. Pouse, R. E. Marvel, R. F. Haglund, D. M. Neumark, and S. R. Leone, Proc. Natl. Acad. Sci. U. S. A. {\bfseries 114}, 9558 (2017).
\bibitem {PM-Paul-science} P. M. Paul, E. S. Toma, P. Breger, G. Mullot, F. Augé, Ph. Balcou, H. G. Muller, and P. Agostini, Science {\bfseries 292}, 1689 (2001).
\bibitem {UKeller-Optica} R. Locher, L. Castiglione, M. Lucchini, M. Greif, L. Gallmann, J. Osterwalder, M. Hengsberger, and U. Keller, Optica {\bfseries 2}, 405 (2015).
\bibitem {Huillier-Science} M. Isinger, R. J. Squibb, D. Busto, S. Zhong, A. Harth, D. Kroon, S. Nandi, C. L. Arnold, M. Miranda, J. M. Dahlström, E. Lindroth, R. Feifel, M. Gisselbrecht, and A. L’Huillier, Science {\bfseries 358}, 893 (2017).
\end{thebibliography}

\end{document}